# Direct imaging of single-molecule electronic states in chemical reactions

Wei Liu[1]†, Meng Li[2]†, Houqiang Teng[3], Heshan Liu[1], Zhi Li[1], Yu Niu[1]*, Ziren Luo[1]*

[1]Center for Gravitational Wave Experiment, Institute of Mechanics, Chinese Academy of Sciences; No. 15 Bei-si-huan West Road, Beijing, 100190, China.

[2]Key Laboratory of Analytical Chemistry for Life Science of Shaanxi Province, School of Chemistry and Chemical Engineering, Shaanxi Normal University, Xi'an, 710062, China

[3]Lanzhou Center for Theoretical Physics and Key Laboratory of Theoretical Physics of Gansu Province, Lanzhou University, Lanzhou, 730000, China.

*Corresponding author. Email: niuyu@imech.ac.cn, luoziren@imech.ac.cn

†These authors contributed equally to this work.

**Chemical reactions involve individual molecules gaining, losing, or sharing electrons[1]. Conventional experimental methods only measure the average behavior of the ensemble[2,3], obscuring probabilistic nature of the reactions[4]. Single-molecule analysis provides insights into the positions[5], pathways[6,7], and dynamics of individual chemical reactions[8,9]. However, it is still a challenge to obtain stochastic information of individual chemical reactions via an imaging scheme. Here, we have developed an optical imaging method that allows us to perform stochastic analysis of individual electrochemical reactions. It bypasses the diffraction limit by enabling the direct imaging of the electronic states of single molecules in an electrochemical reaction. We introduce surface plasmon resonance and electrode potential modulation to reduce the measured electron density background on a macroelectrode surface to the order of 10 μm$^{-2}$, which makes subtle electron density variations from different electronic states of single molecules visible. The technique allows us to measure individual redox reactions, identify different redox states of single molecules in reactions, and more importantly, analyze the reaction probability fluctuations under different reaction conditions. The proposed imaging scheme not only obtains the electron transfer information in a chemical reaction directly but also provides stochastic information of chemical processes, yielding unique insights into single-molecule reactions. We anticipate our imaging technique will facilitate the study of broad communities, including biology, materials, computational science, and other scientific communities.**

## Electronic states imaging principle

Chemical reactions occur by chance. The stochastic aspects of reactions play a prominent role in biology system[10-12] and catching increasing attention of materials[13-16] and computational science communities[17,18]. The probabilistic nature of individual chemical reactions can be analyzed through statistical methods. To observe the stochastic properties of single-molecule chemical reactions, a scheme is required that can sensitively detect the electron transfer information of a single molecule and simultaneously record multiple reactant molecules in the region. Current scattering-based methods can only track the positions of single molecules[19,20], yet providing limited information about the reactions. Additionally, the size of molecules limits the scattering intensity[21-23]. In contrast, scanning probe techniques can achieve atomic-scale precision[24,25] and provide electron transfer information of the molecule[26], but they are limited to confined target locations.

Instead of molecular size, we propose to visualize electron density variations induced from different molecular electronic states to obtain electron transfer information during individual electrochemical reactions. Electrochemical reactions depend on two key electronic states of single molecules: the highest occupied molecular orbital (HOMO) and the lowest unoccupied molecular orbital (LUMO). Electrons can transfer across these molecular orbital levels, with the HOMO holding electrons at the highest energy and the LUMO remaining empty and capable of receiving electrons[27]. During electrochemical reactions, the electrode oxidizes the molecule by removing electrons from the HOMO level, while it reduces the molecule by adding electrons to the LUMO level.

When a bias voltage is applied to an electrode, it can modulate the Fermi level of the surface to align with the molecular orbitals when a molecule adsorbs on the surface. This alignment enables electrons to transfer across the metal-molecule junction, changing surface electron density around the levels (Fig. 1a). However, the enormous electron background on a gold electrode, with a surface electron density of about $10^5$ $\mu m^{-2}$ at room temperature[28], has made it challenging to detect electron density variations due to single-molecule adsorption (Supplementary Information Section 1).

Surface plasmon resonance (SPR) is a phenomenon in which the electric field of an evanescent wave can excite the collective oscillation of surface electrons. This sensitivity to electron density in the metal-electrolyte interface has made SPR a valuable tool for detecting charged particles at sensing surfaces[29] and surface electron density variations resulting from potential modulation[30-32]. However, not all background electrons participate in SPR. Based on our understanding of solid-state physics, only electrons near the Fermi surface can be excited by an applied electric field in metal[33,34]. In momentum space (Fig. 1b), the collective oscillation of the free electrons with the wave vector $k_{sp}$ suggests the electron density involving in SPR, $n_{sp}$, be only a fraction of the background surface electron density, $n_s$. We can determine $n_{sp}$ by

$$n_{sp} = \frac{k_{sp}}{k_F} n_s \tag{1}$$

where $k_F$ is the Fermi wave vector for gold, around $1.21 \times 10^4$ $\mu m^{-1}$ and we can estimate $k_{sp}$ by the SPR condition, the wave vector matching of the surface plasmon and the evanescent wave, that is $k_{sp} = k_x = 2\pi/\lambda \approx 10$ $\mu m^{-1}$ for the visible light with the wavelength $\lambda = 0.63$ $\mu m$. Consequently, the electron density in response to SPR is only a thousandth of the surface electron density.

Eq. 1 confers an extraordinary sensitivity on SPR to the surface electron density variations from the single-molecule interactions in an electrochemical reaction. To exemplify, the measured electron density background can be reduced to the order of 10 $\mu m^{-2}$, or $10^3$ per pixel, when a potential modulation around 100 mV is introduced (Supplementary Information Section 2). In contrast, the electron background in the area amounts to approximately $10^9$. Therefore, to measure electron density variations on the order of $10^{-4}$ per pixel, we introduce a sinusoidal potential to the electrode with an amplitude of 200 mV AC and different DC components in proximity to the HOMO/LUMO level at 1.1 Hz.

The introduction of the potential modulation offers three distinct advantages. Firstly, it can align the Fermi level of the electrode within the electronic states of the target, enhancing the electron transfer between the adsorbed molecule and the surface. Secondly, the response to the potential modulation can also provide an observable, namely the electron density variations around the aligned level. Additionally, the signal-noise-ratio of the subtle signals could be enhanced by the sinusoidal modulation. On the one hand, this approach integrates the electron density variations around the degenerate orbital levels during integration time, one minute in our experiment. On the other hand, it makes signal Fourier analysis applicable which is a powerful tool to analyze subtle signals as illustrated in Extended Figs. 1a to 1d.

Detecting the electron density variations offers an additional advantage: it overcomes the diffraction limit. Instead of focusing on molecular size, we concentrate on the molecular electronic state. The localization accuracy of individual chemical reactions can be relaxed to tens to hundreds of square micrometers. As a result, we can perform multiple reactant molecules measurements within a large region without the need to consider interactions among molecules, which allows the independence and randomness of individual reactions. In our prism-coupled SPR imaging system (Fig. 1c), we can measure several hundred individual reactions spread over a 1 $mm^2$ area in the working cell. During the reactions, we obtain surface information from two images: one in the time domain which records the surface background and the other in the frequency domain from which we can deduce the electron density variations resulting from single-molecule reactions.

**Individual redox reactions measurement**

To demonstrate the efficacy of our imaging scheme, we applied it to the study of the oxidation of methylene blue (MB), a classical redox agent[35]. When the electrode removes an electron from the HOMO level of a MB molecule, the molecule transitions to an energy level one level below the HOMO (HOMO-1) and returns to the HOMO level when an electron is added (Fig. 2a). The cyclic voltammetry of MB exhibited oxidation around 200 mV and reduction around 50 mV (Fig. 2b). By sequentially applying four step potentials (each varying by 200 mV) from -200 mV to 400 mV with sinusoidal modulation, we probed a 10 aM MB modified surface (Fig. 2c). As the cell volume was approximately 50 μL, several hundred redox molecules were present. The time-domain response (DC grayscale) of the entire region in Fig. 2d and the temporal evolution of the corresponding frequency-domain signal (AC grayscale) in Fig. 2e both exhibited stepwise responses. However, while the time-domain response reflected the oscillation around the four potential steps (Extended Data Figs. 2a to 2d and Supplementary Video 1), as suggested in our previous work[30], the frequency-domain signal evolution responded to the electron density variations of the surface around these steps, i.e. the number of MB molecules oxidized around these steps（Extended Data Figs. 2e to 2h and Supplementary Video 2）. As the majority of MB molecules were oxidized around 200 mV, the response peaked and decreased with the oscillation around 400 mV due to decreased efficiency of electron transfer.

**Single-molecule redox states identification**

We need a comprehensive analysis of frequency-domain signals to accurately identify individual redox reactions. Under a potential modulation, frequency-domain signals measure the electron density variations during the integration time (Supplementary Information Section 4). The variations come from two distinct sources: those from thermal motion of electrons on the surface, serving as the background noise, and those arising from single-molecule redox reactions. Therefore, it is crucial to establish guidelines that can distinguish between the sources of signal variations within a pixel.

Electronic thermal noise tends to follow a Gaussian distribution[36], making it possible to minimize its effects through signal averaging or by setting a confidence level of 3 standard deviations of the background noise for detecting oxidations in a pixel. In Extended Data Fig. 3, the blank control pixels in which no MB molecules are immobilized indicate that the background noise reaches an upper limit of 4 grayscale levels. Because the frequency-domain signal in a pixel caused by one electron per minute is approximately 4 grayscale levels (Supplementary Figs. 1a and 1b), at most one electron excited from the background during the integration period. Once the signal surpasses the threshold of 4 grayscale levels, it becomes possible to infer the presence of an individual oxidation in the pixel.

In Fig. 3a, we present the temporal evolution of the normalized frequency-domain signal distribution in the acquired frames. The red dashed line indicates the mean value of thermal noise, which is approximately 2 grayscale levels. Extended Data Fig. 5 demonstrates that the number of pixels within this range remains stable implying that the background thermal noise is barely affected by the step voltage under the four conditions. The blue dashed line represents the upper limit of the confidence level, 4 grayscale levels. In Fig. 3b, we only count the pixels whose signal exceeds this limit. The resulting plot shares a similar shape to that in Fig. 2e, suggesting the step potentials modulate the number of individual electrochemical reactions.

We have plotted histograms showing distributions of the identified oxidation during the -200 mV stage from 0 seconds to 200 seconds in Fig. 3c and the 200 mV stage from 600 seconds to 800 seconds in Fig. 3d. Both histograms follow Poisson distributions, suggesting that the identified oxidations occur randomly and are not dependent on each other. During the -200 mV stage, the identified oxidation has a mean of 1.9, indicating that less than 2 oxidations can be identified per frame. However, the number of identified oxidations significantly increases to 33.4 during the 200 mV stage from 600 seconds to 800 seconds in Fig. 3d. In addition, the histogram shifts to a Gaussian-like distribution with the increased number of identified oxidations.

We can further analyze redox reactions in individual pixels by comparing their signal with that of a blank control pixel. In Fig. 3e, we selected two pixels for analysis: ROI 1 with single-molecule redox reactions and ROI 2 as the blank control. The distributions of the frequency-domain signal evolution for both areas are illustrated in Fig. 3f. While the signal in the blank control exhibited a normal distribution shape, the signal in ROI 1 had a skewed distribution. The frequency of the signal exceeding the mean value of the thermal noise, which was 2 grayscale levels, increased, and half of the signals reached a high amplitude above 4 grayscale levels during the experiment. We can analyze the redox process in the frequency-domain signal evolution diagram depicted in Fig. 3g, which has identified at least 5 redox processes in ROI 1. The signal increased from 2 grayscale levels to the neighborhood of 6 grayscale levels, and the majority of redox processes occurred under 0 mV to 400 mV oscillation, compared with the blank control signal, which remained below 3.6 grayscale levels. Therefore, besides thermal noise, one additional electron caused a change in electron density due to the oxidation of MB molecules.

**Stochastic chemical kinetics analysis**

Different from other single-molecule imaging techniques, we can obtain stochastic information about single-molecule reactions in which broad scientific community are interested, owing to the ability to analyze multiple single-molecule reactions simultaneously at the expense of decreased precision in locating single molecule positions. For example, in ensemble analysis, reaction efficiency is described by the reaction rate, whereas for individual molecules, the probability of single-molecule reaction and the duration of the products determine the stochastic chemical kinetics, both of which can be obtained by statistically analyzing multiple independent single-molecule reactions.

Figs. 4a to 4d compare the mean frequency-domain signals of the pixels in the reaction region under four different step potentials. The mean values of these pixels indicate the average durations of the electron density variations under different potentials, influenced by the oxidation probability of MB molecules. Figs. 4e to 4h show the corresponding grayscale distributions. To estimate the probability fluctuations under four conditions, we fit the grayscale distributions using binomial distribution (Supplementary Information Section 5). As a result, the probabilities of single electrochemical reactions under different conditions increase from 0.16 at -200 mV potential oscillation to 0.21 at 0 mV, peak at 200 mV with 0.34, and then decrease to 0.29 at 400 mV (Fig. 4i). Correspondingly, the maximum expected durations of MB molecule oxidation increase from 6.7 s at -200 mV to 8.5 s at 0 mV, peak at 200 mV with 14.3 s and then decrease to 11.9 s at 400 mV.

**Conclusions and outlooks**

By revisiting SPR theory from a solid-state perspective, we have developed an optical imaging scheme to analyze the stochastic aspects of individual electrochemical reactions. Two prerequisites give the scheme an ability to visualize individual chemical reactions in a relatively large region: the measured electron density background on an electrode surface is reduced to the order of 10 $\mu m^{-2}$ and the alignment between the surface Fermi level and the frontier molecular orbitals enhances the electron transfer between the surface and the single molecules. As a demonstration of this technique, we use the scheme to analyze the oxidation of single MB molecules.

Directly visualizing the electronic states of the individual molecules, this single-molecule imaging method can be applied to a wide range of molecules, especially small molecules difficult to label. Further, it provides unique insights into chemical processes. Compared with the scattering-based methods, it not only bypasses the limitations of the diffraction limit but also provides electron transfer information during a reaction. It also brings stochastic information of single-molecule reactions due to the high-throughput characteristic allowing for multiple single-molecule reaction statistics. Because of the compatibility with microscopy imaging scheme[37], the technique can be developed as a single-molecule SPR microscope when the light source noise is carefully depressed. We believe this method offers an alternative and complementary single-molecule imaging approach that may prove useful for biology, materials, computational science, and other scientific communities.

**References**

1   Marcus, R. A. Electron transfer reactions in chemistry. Theory and experiment. *Reviews of modern physics* **65**, 599 (1993).
2   Bard, A. J. & Faulkner, L. R. *Electrochemical Methods: Fundamentals and Applications, 2nd Edition*. (Wiley, 2000).
3   Zoski, C. G. *Handbook of Electrochemistry*. 877 (Elsevier, 2007).
4   in *The IUPAC Compendium of Chemical Terminology*  Ch. R05176, (2014).
5   Lelek, M. *et al.* Single-molecule localization microscopy. *Nat Rev Methods Primers* **1** (2021). https://doi.org:10.1038/s43586-021-00038-x

6   Liu, S., Bokinsky, G., Walter, N. G. & Zhuang, X. Dissecting the multistep reaction pathway of an RNA enzyme by single-molecule kinetic "fingerprinting". *Proc Natl Acad Sci U S A* **104**, 12634-12639 (2007). https://doi.org/10.1073/pnas.0610597104
7   Xu, W., Kong, J. S., Yeh, Y. T. & Chen, P. Single-molecule nanocatalysis reveals heterogeneous reaction pathways and catalytic dynamics. *Nat Mater* **7**, 992-996 (2008). https://doi.org/10.1038/nmat2319
8   Zhang, Y., Song, P., Fu, Q., Ruan, M. & Xu, W. Single-molecule chemical reaction reveals molecular reaction kinetics and dynamics. *Nat Commun* **5**, 4238 (2014). https://doi.org:10.1038/ncomms5238
9   Guan, J. *et al.* Direct single-molecule dynamic detection of chemical reactions. *Sci Adv* **4**, eaar2177 (2018). https://doi.org:10.1126/sciadv.aar2177
10  Tu, Y. The nonequilibrium mechanism for ultrasensitivity in a biological switch: sensing by Maxwell's demons. *Proc Natl Acad Sci U S A* **105**, 11737-11741 (2008). https://doi.org/10.1073/pnas.0804641105
11  Fange, D., Berg, O. G., Sjoberg, P. & Elf, J. Stochastic reaction-diffusion kinetics in the microscopic limit. *Proc Natl Acad Sci U S A* **107**, 19820-19825 (2010). https://doi.org/10.1073/pnas.1006565107
12  Qian, H. Phosphorylation energy hypothesis: open chemical systems and their biological functions. *Annu Rev Phys Chem* **58**, 113-142 (2007). https://doi.org/10.1146/annurev.physchem.58.032806.104550
13  Longo, S., Micca Longo, G., Hassouni, K., Michau, A. & Prasanna, S. Stochastic models of systems for Nanotechnology: from micro to macro scale. *Nanotechnology* **32**, 145604 (2021). https://doi.org:10.1088/1361-6528/abd2ea
14  Gupta, K. K., Roy, L. & Dey, S. Hybrid machine-learning-assisted stochastic nano-indentation behaviour of twisted bilayer graphene. *Journal of Physics and Chemistry of Solids* **167**, 110711 (2022). https://doi.org:https://doi.org/10.1016/j.jpcs.2022.110711
15  Valle, J. D. *et al.* Generation of Tunable Stochastic Sequences Using the Insulator-Metal Transition. *Nano Lett* **22**, 1251-1256 (2022). https://doi.org:10.1021/acs.nanolett.1c04404
16  Park, T. J. *et al.* Efficient Probabilistic Computing with Stochastic Perovskite Nickelates. *Nano Letters* **22**, 8654-8661 (2022). https://doi.org:10.1021/acs.nanolett.2c03223
17  Alaghi, A. & Hayes, J. P. Survey of stochastic computing. *ACM Transactions on Embedded computing systems (TECS)* **12**, 1-19 (2013).
18  Kari, S. R. Principles of Stochastic Computing: Fundamental Concepts and Applications. *arXiv preprint arXiv:2011.05153* (2020).
19  Kukura, P. *et al.* High-speed nanoscopic tracking of the position and orientation of a single virus. *Nat Methods* **6**, 923-927 (2009). https://doi.org:10.1038/nmeth.1395
20  Ortega Arroyo, J. *et al.* Label-free, all-optical detection, imaging, and tracking of a single protein. *Nano Lett* **14**, 2065-2070 (2014). https://doi.org:10.1021/nl500234t
21  Arroyo, J. O. & Kukura, P. Non-fluorescent schemes for single-molecule detection, imaging and spectroscopy. *Nature Photonics* **10**, 11-17 (2016). https://doi.org:10.1038/nphoton.2015.251
22  Kneipp, K. *et al.* Single Molecule Detection Using Surface-Enhanced Raman Scattering (SERS). *Physical Review Letters* **78**, 1667-1670 (1997). https://doi.org:10.1103/PhysRevLett.78.1667
23  Nie, S. Probing Single Molecules and Single Nanoparticles by Surface-Enhanced Raman Scattering. *Science* **275**, 1102-1106 (1997). https://doi.org:10.1126/science.275.5303.1102
24  Binnig, G. & Rohrer, H. Scanning tunneling microscopy. *Surf. Sci.* **126**, 236-244 (1983).

25    Binnig, G., Quate, C. F. & Gerber, C. Atomic force microscope. *Physical review letters* **56**, 930 (1986).
26    de Oteyza, D. G. *et al.* Direct imaging of covalent bond structure in single-molecule chemical reactions. *Science* **340**, 1434-1437 (2013).
27    Baldo, M. *Introduction to nanoelectronics*.  (MIT OpenCourseWare, 2011).
28    Bürgi, L., Knorr, N., Brune, H., Schneider, M. & Kern, K. Two-dimensional electron gas at noble-metal surfaces. *Applied Physics A* **75**, 141-145 (2002).
29    Shan, X. *et al.* Measuring surface charge density and particle height using surface plasmon resonance technique. *Anal Chem* **82**, 234-240 (2010). https://doi.org/10.1021/ac901816z
30    Liu, W., Niu, Y., Viana, A. S., Correia, J. P. & Jin, G. Potential Modulation on Total Internal Reflection Ellipsometry. *Analytical Chemistry* (2016).
31    Abayzeed, S. A., Smith, R. J., Webb, K. F., Somekh, M. G. & See, C. W. Sensitive detection of voltage transients using differential intensity surface plasmon resonance system. *Opt Express* **25**, 31552-31567 (2017). https://doi.org:10.1364/OE.25.031552
32    Foley, K. J., Shan, X. & Tao, N. J. Surface impedance imaging technique. *Analytical Chemistry* **80**, 5146-5151 (2008). https://doi.org:10.1021/ac800361p
33    Ziman, J. M. Electrons in metals: A short guide to the fermi surface. *Contemporary Physics* **4**, 81-99 (1962).
34    Snoke, D. W. *Solid state physics: Essential concepts*.  (Cambridge University Press, 2020).
35    Clark, W. M., Cohen, B. & Gibbs, H. D. Studies on Oxidation-Reduction: VIII. Methylene Blue. *Public Health Reports (1896-1970)* **40** (1925). https://doi.org/10.2307/4577559
36    Price, J. & Goble, T. in *Telecommunications Engineer's Reference Book*   (ed Fraidoon Mazda)  10-11-10-15 (Butterworth-Heinemann, 1993).
37    Shan, X., Patel, U., Wang, S., Iglesias, R. & Tao, N. Imaging local electrochemical current via surface plasmon resonance. *Science* **327**, 1363-1366 (2010). https://doi.org:10.1126/science.1186476

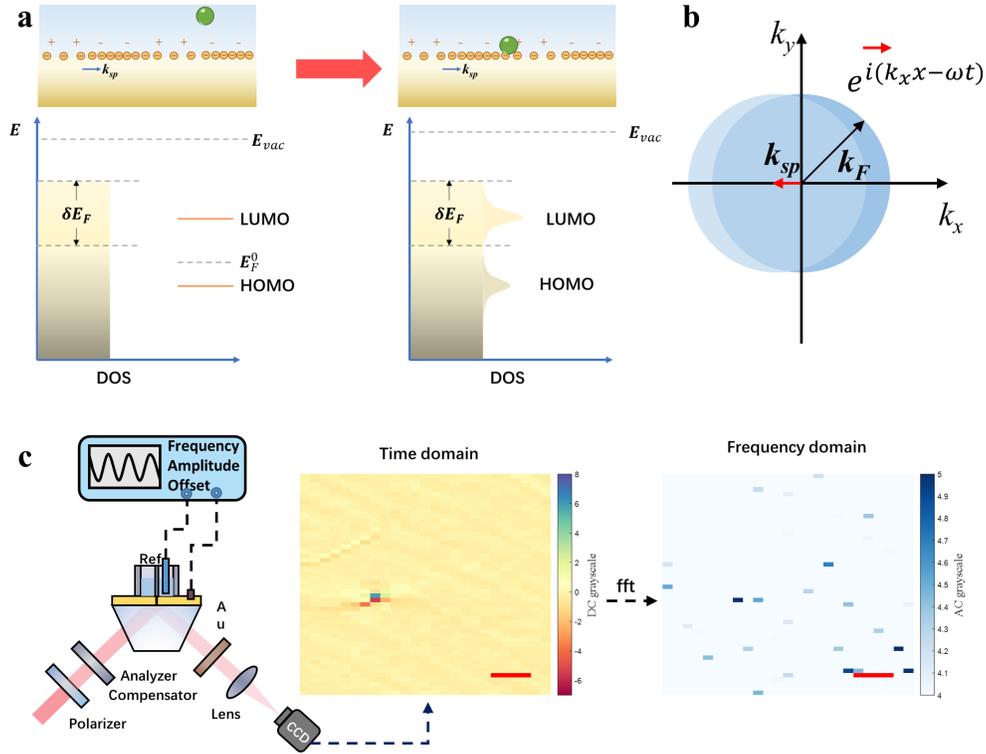

**Fig. 1 | Electronic states imaging principle and optical setup. a.** When the Fermi level of the electrode is aligned with the degenerate molecular orbital level after the adsorption of an individual molecule, such as the highest occupied molecule orbital (HOMO) or the lowest unoccupied molecular orbital (LUMO), the electron density around the level will change because the electron transfer changes the density of states (DOS, the number of different states at a particular energy level that electrons are allowed to occupy) within the vicinity of the level. $E_{vac}$ is the vacuum energy level, which is identical for an isolated metal surface and an isolated molecule. $E_F^0$ is the Fermi level of the molecule and $\delta E_F$ the Fermi level variation modulated by a bias potential applied to the surface. **b.** In momentum space, the collective oscillation of a two-dimensional electron gas with the wave vector $k_{sp}$ matching the wave vector of a unit evanescent wave, $k_x$. Only electrons near the Fermi surface participate in the oscillation. **c.** The optical setup of the prism-based SPR imaging system used to measure the single-molecule electronic states. We can obtain two images at the same time: one in the time domain, and the other in the frequency domain. The scale bar is 100 $\mu$m.

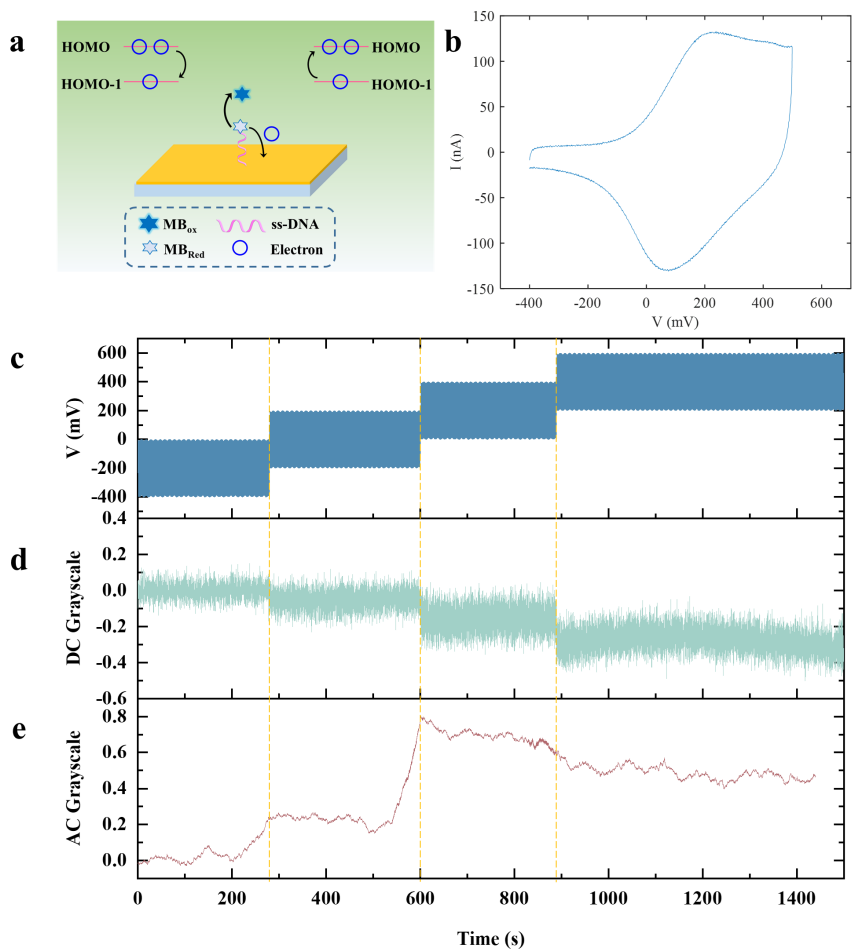

**Fig. 2 | Time-domain signal and Frequency-domain signal evolution of 10 aM methylene blue (MB) oxidation under different potentials. a.** The oxidative–reductive mechanism of MB. **b.** Cyclic voltammetry of 1 $\mu$M immobilized MB in a buffer. **c.** Four-step potential oscillations: -400 mV to 0 mV, -200 mV to 200 mV, 0 mV to 400 mV, and 200 mV to 600 mV. **d.** The synchronized time-domain signal. **e.** The synchronized frequency-domain signal evolution. Because each point of the curve was the one-minute integration of the time-domain signal, the frequency-domain signal was advanced by 60 seconds compared to the time-domain signal.

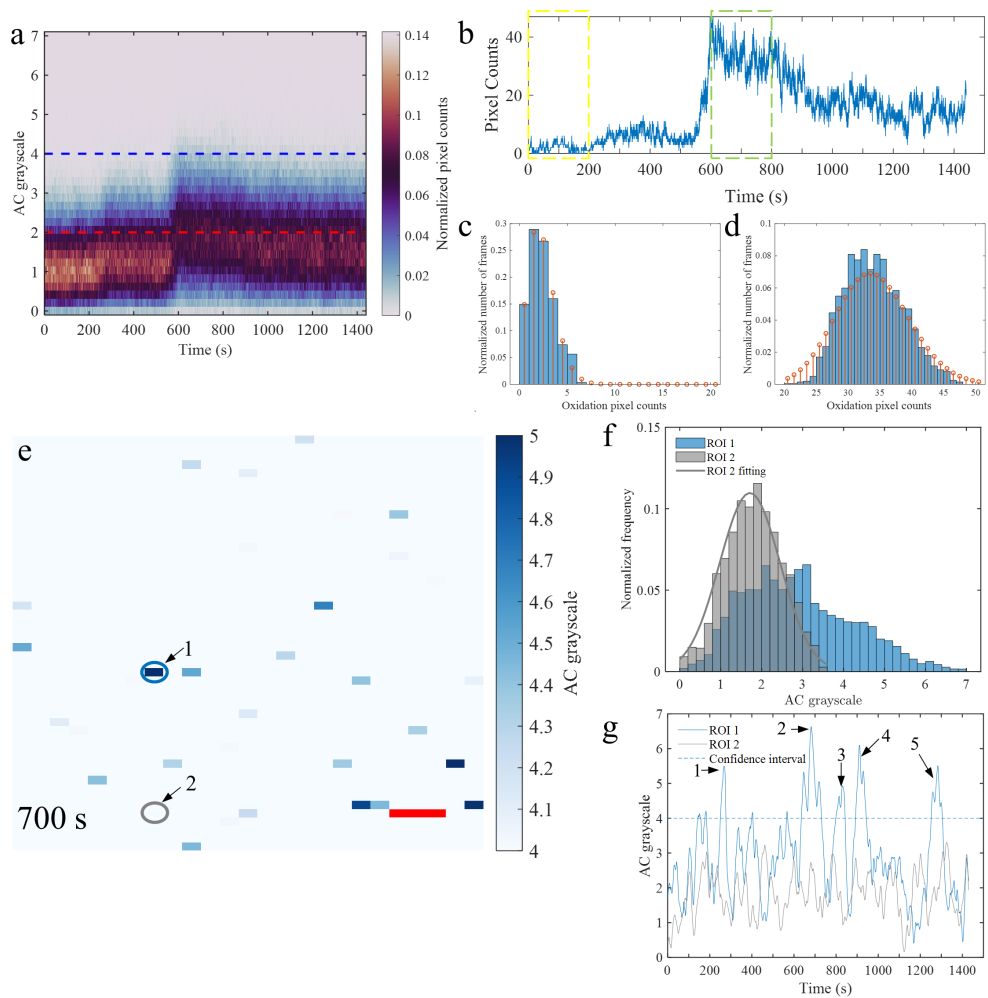

**Fig. 3 | Identification of individual oxidation of MB molecules. a.** The evolution of frequency-domain signal distribution. The distribution of frequency-domain signal at a specific time can be obtained by the cross-sectional view of the figure along the time. For example, Extended Data Figs. 5a to 5d depict the distributions of the frequency-domain signals at 0 s, 400 s, 700s and 1000 s. The red dash line indicates the mean value of thermal noise and the blue dash line the upper limit of thermal noise. **b.** The evolution of the number of pixels the intensity of which exceeds the upper limit of thermal noise, 4 grayscale levels. Two stages are selected: 0 s to 200 s and 600 s to 800 s. **c.** The histogram of oxidation pixels per frame during 0 s to 200 s. **d.** The histogram of oxidation pixels per frame during 600 s to 800 s. **e.** The frequency-domain frame at 700 s. Two pixels are selected: individual oxidation of a MB molecule is identified in ROI 1 and ROI 2 as a blank control pixel. **f.** The frequency-domain signal distribution of the two selected regions during the experiment. **g.** The evolution of frequency-domain signal of the two selected regions during the experiment.

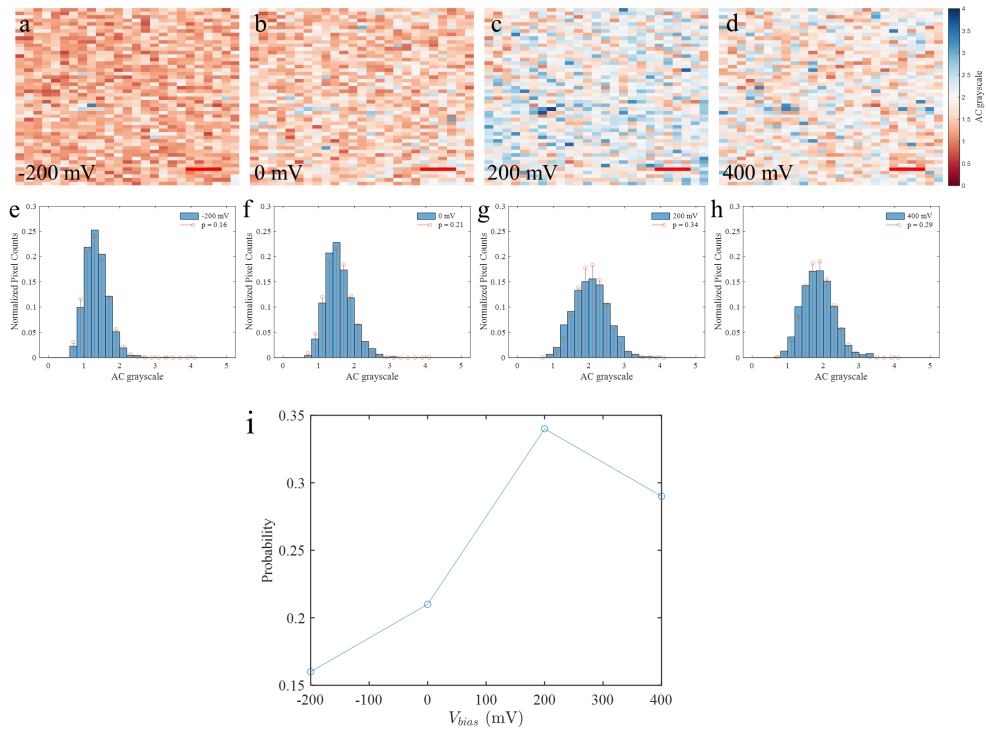

**Fig. 4 | Reaction probability analysis of individual MB oxidations. a**, **b**, **c** and **d**, The average frames under the four step potentials. Each frame is the average of 2001 sequential frames, i.e., 0 s ~ 200 s for **a**, 300 s ~ 500 s for **b**, 600 s ~ 800 s for **c**, and 900 s ~ 1100 s for **d**. The color bar indicates the intensity of the frequency-domain signals, AC grayscale. The scale bar was 100 $\mu$m. **e**, **f**, **g**, and **h**, The histograms of the frequency-domain signals under the four conditions and the corresponding binomial distribution fitting. **i**. The reaction probability under the four potentials.

# Methods

## Sample preparation
12-base single-strand DNA (GCGCGGCGCGCG) modified with a sulfhydryl group at the 3-end and a MB group at the 5-end and Tris-(2-carboxyethyl)-phosphine hydrochloride (TCEP), and Tris-EDTA buffer were purchased from Sangon Biotech (Shanghai) Co., Ltd. phosphate buffer (PB) from J&K Scientific.

Prior to immobilization of MB-labeled aptamer onto the gold surface, 100 μM aptamer was dissolved in Tris-EDTA buffer solution with 1 mM TCEP and incubated in the dark for 2 h to reduce the disulfide bonds. Then, the aptamer was diluted to 10 aM and injected into the working cell by a pump, allowed to stay for 1-2 h to self-assembly the MB-labeled aptamer on the surface of the gold. The resulted surface was thoroughly washed with 10 mM PB to remove the unbound aptamer.

## Electrochemical measurement
Cyclic voltammetry and open circuits recording step potential oscillations are performed using a VersaSTAT 3 electrochemical system (Princeton, U.S.A). The sensing surface as the working electrode and a Platinum wire as the counter.

## Optical setup
Our prism coupled phase-sensitive SPRi system combines SPR phenomenon with a polarizer-compensator-sample-analyzer (PCSA) imaging ellipsometry configuration. It uses a 625 nm LED as the light source, a high-speed cool CCD camera (Andor iKon-M)as the detector, and a SF10 trapezoidal prism to couple the light and a 48 nm gold-covered sensing surface sensing surface. The light is guided by an optical fiber and expanded by a collimating system. After passing a polarizer and a 45° compensator, the polarized collimated beam propagates perpendicularly to the prism and onto the sensing surface. The incident angle is optimized at 58°, in the neighborhood of ellipsometric singularity of the system where the evanescent wave appears at the sensing surface to detect the interaction in very shallow depth from the surface and the phase of the p-polarized reflection jumps. The reflected light carrying the surface information is then imaged by a CCD camera at a frame rate of 10 Hz after passing an analyzer. To achieve the best sensitivity, the system works under optimized linear off-null mode[38] where the azimuth of the polarizer is optimized at 90° and that of the analyzer at 45°. During the measurement, the optical signal variation from the potential modulation is recorded by CCD in grayscale.

The sensing chip is divided into two adjacent electric insulated areas: one as the working cell connecting to a signal generator (Agilent 33210A), the other as the reference cell. In the working cell, In the working cell, the gold-covered sensing surface is the working electrode and a Pt wire is the counter.

## Data analysis
**Power fluctuation depression and amplitude density spectrum analysis.** In order to measure electron density variations on the order of $10^{-4}$ per pixel, we need to depress light power fluctuations during the experiment at first. The large field of view of the imaging scheme can allow for the simultaneous recording of reflections from both the working cell and the reference cell. Throughout an experiment, a sequence of images is captured (Extended Data Fig. 1a), and two regions of interest (ROI) are selected: ROI 1 in the working cell and ROI 2 in the reference cell. Because the reflections from both regions experience similar fluctuations in light source power during the measurement as shown in Extended Data Fig. 1b, the differential signal between the two regions can significantly reduce the effects of these fluctuations. To further improve the signal-to-noise ratio, a sinusoidal potential modulation with an amplitude of 200 mV AC and different DC components at 1.1 Hz is introduced to the working cell.

The comparison between the recorded signal in ROI 1 and the differential signal in Extended Data Fig. 1c demonstrates a significant improvement in the signal-to-noise ratio. The power fluctuations in the recorded signal are measured to be $4 \times 10^{-3}$, whereas the differential signal fluctuates less than $10^{-3}$. The modulation signal is barely noticeable in the raw signal but can be clearly distinguished in the differential signal. The corresponding amplitude density spectrum in Extended Fig. 1d provides a comprehensive noise analysis. Prior to the depression scheme, the noise ranges from $10^{-3}$ Hz$^{-1/2}$ within the frequency band from 10 Hz to 1 Hz, and approaches $10^{-2}$ Hz$^{-1/2}$ within the low-frequency band from 1 Hz to $10^{-2}$ Hz. After the differential, the noise is near the signal level, $10^{-4}$ Hz$^{-1/2}$, while the integration of the modulated signal at 1.1 Hz for one minute amplifies the measured signal to above $10^{-3}$ Hz$^{-1/2}$, providing the technique with an ability to analyze single-molecules electronic states.

**Time-domain frame averaging.** To improve the signal-to-noise ratio of time-domain frames, we take the average frame at the time to eliminate thermal noise. In detail, at $t$ seconds, a sequential of 600 frames was recorded. The average of these frames was taken as the frame at $t$ seconds.

**Normalization of statistical data.** There are 1250 pixels in a frame for our configuration. For a given frame at a certain time, the pixels are categorized by their intensities, AC grayscale. Each intensity interval is 0.2 grayscale levels. The normalization of pixel counts in Figs. 3a, 4e to 4h and Extended Data Fig. 5 is dividing the number of pixels the intensity of which falls into each interval by the total number of the pixels, 1250.

In Figs. 3c and 3d, we count the number of oxidation pixels the intensity of which exceeds the upper limit of thermal noise, 4 grayscale levels in 2001 consecutive frames, 0 s ~ 200 s for Fig. 3c and 600 s to 800 s for Fig. 3d. The frames are categorized by the number of the spotted oxidation pixels. The normalization of the number of frames is dividing the number of frames enjoying the same number of the oxidation pixels by the number of consecutive frames, 2001.

For individual pixels, we categorize their intensities. Each intensity interval is 0.2 grayscale levels. We get the normalized frequency in Fig. 3f by dividing the frequency of each intensity category by the total number of frames, 14400.

**The average frame under the four step potential oscillations.** The frames in Figs. 4a to 4d are the average frame of 2001 consecutive frames under the four step potential oscillations: 0 s ~ 200 s for Fig. 4a, 300 s to 500 s for Fig. 4b, 600 s to 800 s for Fig. 4c, and 900 s to 1100 s for Fig. 4d.

**The probability fitting.** We use least squares method to fit the binomial distribution in Figs. 4e to 4h.

**Data availability**

All data that support the findings of this study are available from the corresponding author upon reasonable request.

38      Jin, X. *et al.* in *4th Optics Young Scientist Summit (OYSS 2020)*   (2021).

**Acknowledgments:** The authors acknowledge financial support from National Key R&D Program of China (2020YFC2200100 and 2021YFC2202902), Director Fund of Institute of Mechanics, and Fundamental Research Funds for the Central Universities (GK202207016). We are grateful to Dr. Sixing Xu for his insightful suggestion that the detection response be from the electron density variation at the very beginning of this work, Miss Yu Miao for her constant urging the author to polish this work.

**Author contributions:** WL, YN, ZRL conceived the idea, WL, YN, ZRL, ML proposed the methodology, WL, ML, ZL carried out the experiment, WL, ML, HT analyzed the data, WL, HL developed the algorithms and YN, ZRL supervised the project. All authors discussed the results and contributed to the writing the manuscript.

**Competing interests:** Authors declare that they have no competing interests.

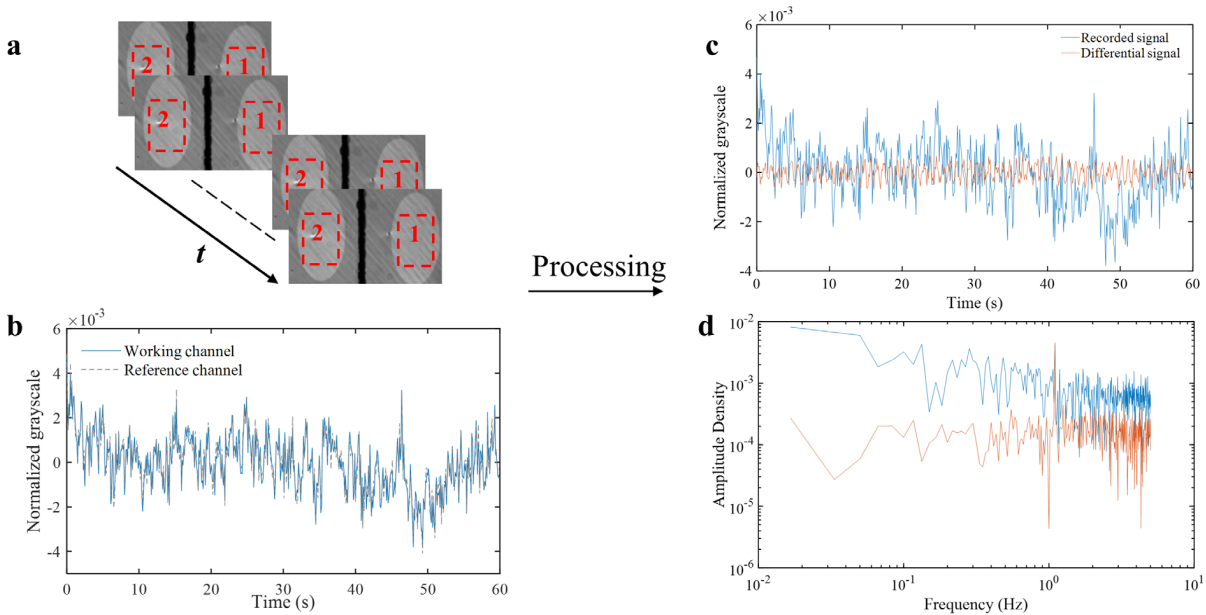

**Extended Data Fig. 1 | The signal process for captured frames. a.** Raw time sequence of captured frames. Two regions of interest (ROI) are selected: ROI 1 in the working cell and ROI 2 in the reference cell. **b.** The raw signals of ROI 1 and ROI 2. **c.** The recorded signal of ROI 1 and the differential signal between ROI 1 and ROI 2 during 60 seconds in the time domain. **d.** The amplitude spectrum density of the raw signal of ROI 1 and the differential signal in the frequency domain.

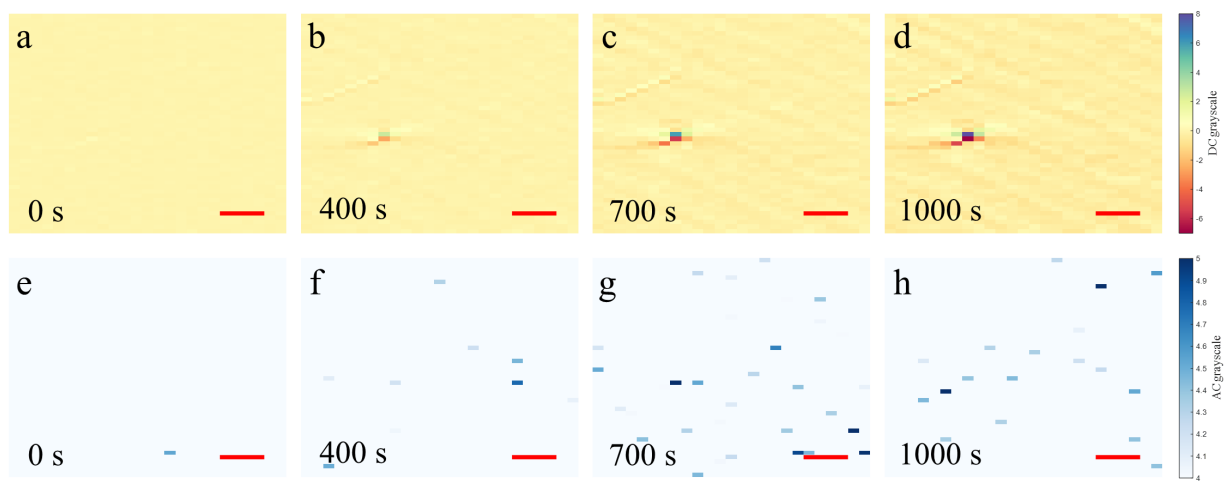

**Extended Data Fig. 2 | Time-domain frames and Frequency-domain frames. a, b, c,** and **d,** Time-domain frames at 0 s, 400 s, 700 s, and 1000 s. Each frame was averaged by a sequential of 600 raw frames at the time to eliminate thermal noise. Time-domain images provide background information about the surface, such as topography, defects, or the solution background. As the step potential increased, scratches and defects became apparent due to the charge distribution, as discussed by Xiaonan Shan[37]. **e, f, g,** and **h,** Frequency-domain frames at the corresponding time. Each frequency-domain frame was integrated for 1 minute. Only pixels the intensity of which exceed thermal noise, 4 grayscale levels, were spotted. During the potential oscillation from -400 mV to 0 mV, a single pixel with oxidized MB molecules was identified at the start of the experiment. As the step potential increased, the pixel counts increased from 9 at 400 s (oscillation from -200 mV to 200 mV) to 32 at 700 s (oscillation from 0 mV to 400 mV). However, instead of following a continuous increase, the number of pixels decreased to 20 at 1000 s (oscillation from 200 mV to 600 mV) due to a decrease in electron transport efficiency.

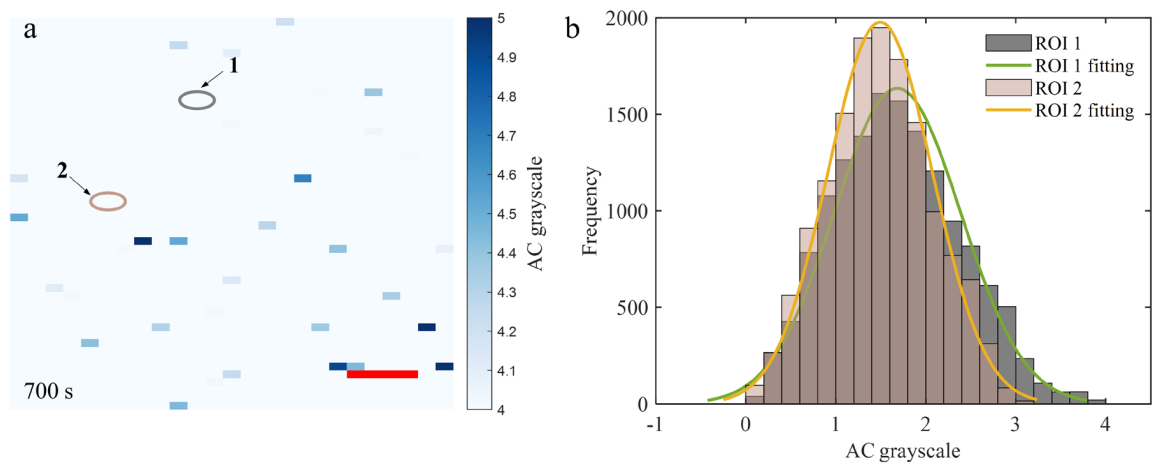

**Extended Data Fig. 3 | Frequency-domain signal distribution of blank control pixels throughout the experiment. a.** Frequency-domain frame at 700 s. Two pixels are selected as blank controls where no MB molecules were immobilized. The scale bar is 100 $\mu$m. **b.** Distribution of Frequency-domain signals of both regions during the experiment. Both regions shared normal distributions because the signals came from thermal noise, the intensity of which were less than 4 grayscale levels, suggesting at most one electron were excited by temperature during one minute.

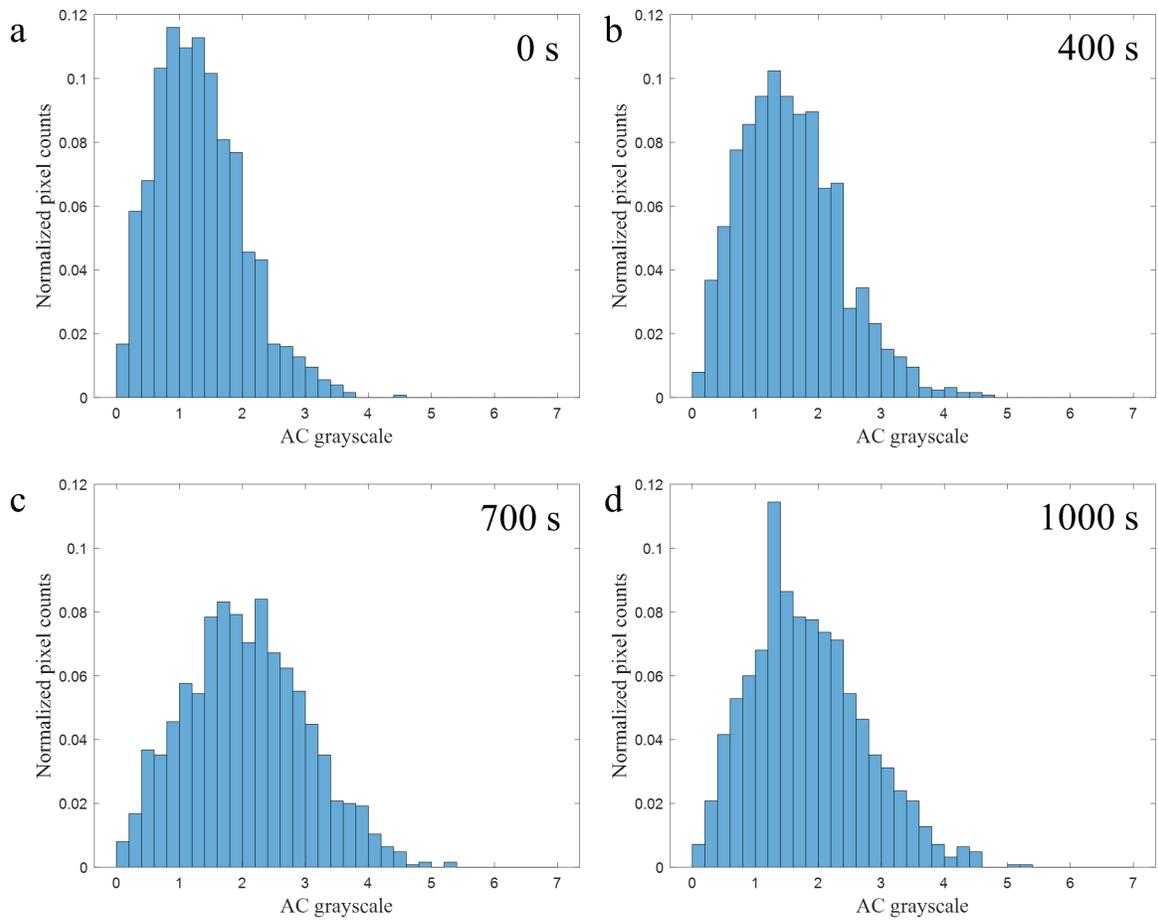

**Extended Data Fig. 4 | Frequency-domain signal distribution at a certain time. a, b, c** and **d.** The histograms of frequency-domain signals at 0 s, 400 s, 700 s and 1000 s.

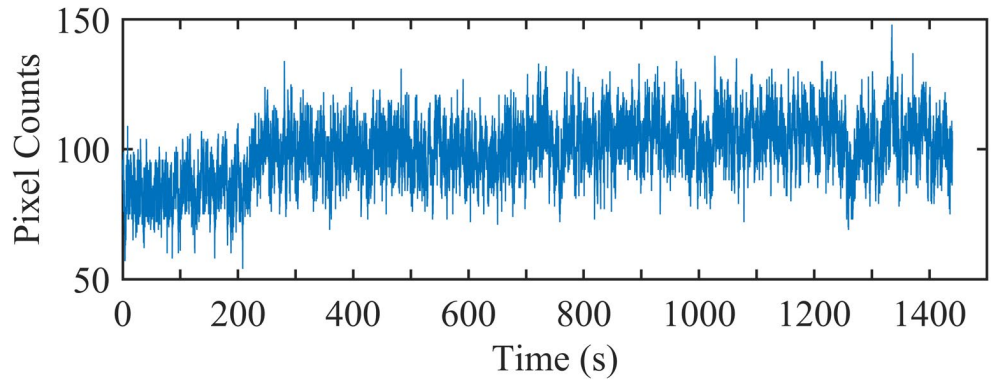

**Extended Data Fig. 5 | The evolution of the number of pixels the intensity between 1.8 to 2.0 grayscale levels.**

# Direct imaging of single-molecule quantum states in chemical reactions

Wei Liu[1]†, Meng Li[2]†, Houqiang Teng[3], Heshan Liu[1], Zhi Li[1], Yu Niu[1]\*, Ziren Luo[1]\*

[1]Center for Gravitational Wave Experiment, Institute of Mechanics, Chinese Academy of Sciences; No. 15 Bei-si-huan West Road, Beijing, 100190, China.

[2]Key Laboratory of Analytical Chemistry for Life Science of Shaanxi Province, School of Chemistry and Chemical Engineering, Shaanxi Normal University, Xi'an, 710062, China

[3]Lanzhou Center for Theoretical Physics and Key Laboratory of Theoretical Physics of Gansu Province, Lanzhou University, Lanzhou, Gansu 730000, China.

\*Corresponding author. Email: niuyu@imech.ac.cn, luoziren@imech.ac.cn

†These authors contributed equally to this work.

## Contents

## Electron background estimation

A two-dimensional electron gas (2DEG) is confined on the surface of the gold[1]. In momentum space, these electrons lie within the Fermi circle where $\mathrm{Re}(k_z) = 0$. For a 2D free electron system, the density of states (DOS) is a constant[2]:

$$D = \frac{m}{\pi \hbar^2} \qquad (S.1)$$

where $m$ is the electron mass, 0.511 MeV and D is about $4 \times 10^6$ eV$^{-1}$ μm$^{-2}$. During the equilibrium of the system at $T=0$ K, the electron density $n_{2D}$ is given by

$$n_{2D} = \frac{k_F^2}{4\pi} \qquad (S.2)$$

where $k_F$ is the Fermi wave vector, for gold $k_F = 1.21 \times 10^4$ μm$^{-1}$ and $n_{2D}$ =$1.17 \times 10^7$ μm$^{-2}$ for gold.

For $T > 0$ K, the free electrons follow Fermi-Dirac statistics in equilibrium by

$$f(E) = \frac{1}{\exp\left(\dfrac{E - E_f}{k_B T}\right) + 1} \qquad (S.3)$$

where $k_B$ is Boltzmann constant. The electron density $n_{2D}$ is given by

$$n_{2D} = \int_0^\infty D f(E) dE = \frac{m k_B T}{\pi \hbar^2} \ln\left(1 + e^{\frac{E_f}{k_B T}}\right) \qquad (S.4)$$

For $T= 298$ K, $k_B T$ is 25.7 meV and $n_{2D}$ is estimated by $\frac{m E_f}{\pi \hbar^2} \approx 4 \times 10^5$ μm$^{-2}$ at $E_f = 0.1$ eV

## Electrons involved in SPR under a potential modulation

Under a potential modulation, $\delta U$, the electrode surface undergoes charging/discharging process. The electron density of the surface, $n_s$, can be estimated by

$$n_s = \frac{c \delta U}{e} \qquad (S.5)$$

where $c$ is the interfacial capacitance density (capacitance per unit area). For gold electrode, the order of the interfacial capacitance density[3] ranges from 1 μF cm$^{-2}$ to 10 μF cm$^{-2}$ and the corresponding electron density is $10^3$ μm$^{-2}$ to $10^4$ μm$^{-2}$ approximately when a potential modulation with 100 mV is introduced. Thus, according to eqs. (1) and (S.5), the order of the electrons involved in SPR, $\delta n_{sp}$, is estimated by 10 μm$^{-2}$, and the order of the measured electron in a pixel, $\rho$, is given by

$$\rho = A\delta n_{sp} \approx 10^3 \quad (S.6)$$

where $A$ is the area which a pixel is covering, 300 $\mu$m² approximately (25 $\mu$m × 12 $\mu$m).

## Measuring electron density variations by phase-sensitive SPR

Our phase sensitive SPR imaging system combines SPR phenomenon with a polarizer-compensator-sample-analyzer (PCSA) imaging ellipsometry configuration. The normalized detected signal $\delta I/I$ can be given by[4]

$$\frac{\delta I}{I} = \frac{\delta \boldsymbol{R}_s}{\boldsymbol{R}_s} + \alpha_1 \delta\psi + \alpha_2 \delta\Delta \quad (S.7a)$$

where $\boldsymbol{R}_s$ is the reflectance of the surface for s-polarized light, and $\alpha_1$ and $\alpha_2$ are the coefficients for the ellipsometric parameter variations, $\delta\psi$ and $\delta\Delta$. Because s-polarized light is insensitive to the surface variation, the first term of the right hand of eq. (S.8a) can be neglected and eq. (S.8a) can be expressed as

$$\frac{\delta I}{I} \approx \alpha_1 \delta\psi + \alpha_2 \delta\Delta \quad (S.7b)$$

The explicit expressions of $\alpha_1$ and $\alpha_2$ are given by

$$\begin{aligned} \alpha_1 &= \frac{2\left[\tan\bar\psi(1+\cos 2A) - \sin 2A \sin(2P+\overline{\Delta})\right]}{\left[1 - \cos 2\bar\psi \cos 2A + \sin 2\bar\psi \sin(2P+\overline{\Delta})\sin 2A\right]} \\ \alpha_2 &= \frac{-\sin 2\bar\psi \cos(2P+\overline{\Delta})\sin 2A}{\left[1 - \cos 2\bar\psi \cos 2A + \sin 2\bar\psi \sin(2P+\overline{\Delta})\sin 2A\right]} \end{aligned} \quad (S.8a)$$

where $\bar\psi$ and $\overline{\Delta}$ are unperturbed values of the ellipsometric parameters $\psi$ and $\Delta$ and $P$ and $A$ are the polarizer and analyzer azimuths respectively. To detect the subtle electron density change, the phase SPR works under the linear off null condition[5], where $A$ is optimized at 45° and P at 90°. Thus, eq. (S. 9a) can be simplified as

$$\begin{aligned} \alpha_1 &= \frac{2\left[\tan\bar\psi + \sin\overline{\Delta}\right]}{1 - \sin 2\bar\psi \sin\overline{\Delta}} \\ \alpha_2 &= \frac{\sin 2\bar\psi \cos\overline{\Delta}}{1 - \sin 2\bar\psi \sin\overline{\Delta}} \end{aligned} \quad (S.8b)$$

On the one hand, according to the Drude theory of metals, the electron density variations in a pixel, $\delta\rho$, will exert influence on the dielectric constant of gold electrode, $\varepsilon$, by

$$\delta\varepsilon = (\varepsilon - 1)\frac{\delta\rho}{\rho} \qquad (S.9a)$$

or

$$\frac{\delta\varepsilon}{\varepsilon} = \left(1 - \frac{1}{\varepsilon}\right)\frac{\delta\rho}{\rho} \qquad (S.9b)$$

where $\delta\varepsilon$ is the dielectric constant variation of gold. According to eq. (S.7), the SPR measured electron density variations in a pixel is $10^{-4}$ approximately. Therefore, the ellipsometric phase variation can be given by

$$\begin{aligned}\delta\psi &= \left(\frac{1}{2}\sin 2\psi\right)\mathrm{Re}(\kappa\delta\varepsilon) = \left(\frac{1}{2}\sin 2\psi\right)\mathrm{Re}\big(\kappa(\varepsilon - 1)\big)\frac{\delta\rho}{\rho}\\ \delta\Delta &= \mathrm{Im}(\kappa\delta\varepsilon) = \mathrm{Im}\big(\kappa(\varepsilon - 1)\big)\frac{\delta\rho}{\rho}\end{aligned} \qquad (S.10)$$

where $\kappa$ demonstrates the extent of reflection polarization modulation by the tiny variation in the dielectric constant of the metal and the explicit expression can be found in our previous work[6].

**Frequency-domain signal analysis**

According to eqs. (S.8b) and (S.11), the maximum detected signal $\delta I_{max}$ in a pixel can be used to measure electron density variations by

$$\delta I_{max} = K\frac{\delta\rho}{\rho} \qquad (S.11)$$

where $K = I \times [\alpha_1(\frac{1}{2}\sin 2\psi)\mathrm{Re}\big(\kappa(\varepsilon - 1)\big) + \alpha_2\mathrm{Im}\big(\kappa(\varepsilon - 1)\big)]$ and $I$ is the background intensity of the surface, 157 grayscale levels during the experiment.

Our frequency-domain signal $I_{AC}$ can be estimated by

$$I_{AC} = \delta I_{max}\sqrt{\frac{\tau f}{2}} \qquad (S.12a)$$

or

$$I_{AC} = K\frac{\delta\rho}{\rho}\sqrt{\frac{\tau f}{2}} \qquad (S.12b)$$

where $\tau$ is the duration of electron density variations during the integration period and $f$ is sampling frequency of CCD, 10 Hz in our configuration. According to eq. (S13b), both electron density variations and corresponding duration determine frequency-domain signals. For typical electron density variations from $10^{-5}$ per pixel per minute to $10^{-2}$ per pixel per minute, the frequency-domain signals change linearly from 0.1 grayscale levels to 100 grayscale levels (Supplementary Fig. 1a). For one electron variation in a pixel fitted in our configuration, 1/2500 per pixel approximately, the frequency-domain signal during one-minute reaches 4 grayscale levels. In Supplementary Fig. 1b, the signal approaches to 4 grayscale levels by square root of the duration time, $\sqrt{\tau}$.

After the modulation, the electron density variations, $\delta\rho$, can be divided into two parts: those from thermal motion of electrons, $\delta\rho_T$, and those from the oxidation of single-molecule oxidation, $\delta\rho_{ox}$, that is,

$$\delta\rho = \delta\rho_T + \delta\rho_{ox} \qquad (S.13)$$

Because the oxidation of MB is a two-level system with the two basis states: the HOMO level, denoted as $|0\rangle$, and the HOMO-1 level, denoted as $|1\rangle$, in general the electron density variations of the system can be expressed as

$$\delta\rho_{ox} = p_0 \delta\rho_0 + p_1\ \delta\rho_1 \qquad (S.14a)$$

where $\delta\rho_0$ and $\delta\rho_1$ are the electron density variations of the two basis states, $p_0$ and $p_1$ the probabilities of finding the system in the two states, respectively. Since $\delta\rho_0 = 0$, eq. (S.15a) can be simplified as

$$\delta\rho_{ox} = p_1 \delta\rho_1 \qquad (S.14b)$$

**Distribution of frequency-domain signals**

The oxidation of a molecule can be taken as a Bernoulli experiment: the molecule is oxidized or not. Therefore, the single-molecule redox should follow a binomial distribution (a discrete probability distribution which gives the probability of getting a certain number of successes in a fixed number of trials, where the outcome of each trial is either success or failure) to some extent. However, instead of the discrete number of successes, the distribution is categorized by the continuous frequency-time signal, which can be interpreted as the duration of the electron density variations and the fixed number of trials as the integration period. As a result, the distribution of the frequency-domain signals under a certain condition describes the probability of different durations of the electron density variations during the integration period:

$$I_{AC} \sim B(n_{AC}, p) \qquad (S.15)$$

where $p$ is the electron density variation probability under the certain condition and $n_{AC}$ the corresponding integral for the integration time, given by:

$$n_{AC} = k, \quad 0.6 + 0.2k < I_{AC} \leq 0.8 + 0.2k, \quad (S16)$$

in which, $k = 0, 1, 2, …, 17$.

According to eq. (S16), the expected frequency signal under the condition fulfills:

$$0.6 + 0.2 \times n_{AC}p < E(I_{AC}) \leq 0.8 + 0.2 \times n_{AC}p \quad (S.17)$$

Because the electron density variations from the single-molecule oxidation change under different potentials while those from the thermal noise remain stable, we take $p$ as an indicator of reaction probability and corresponding time for $E(I_{AC})$ as the expected oxidation duration under the condition, which can be obtained from Supplementary Fig. 1b.

## Supplementary Figures

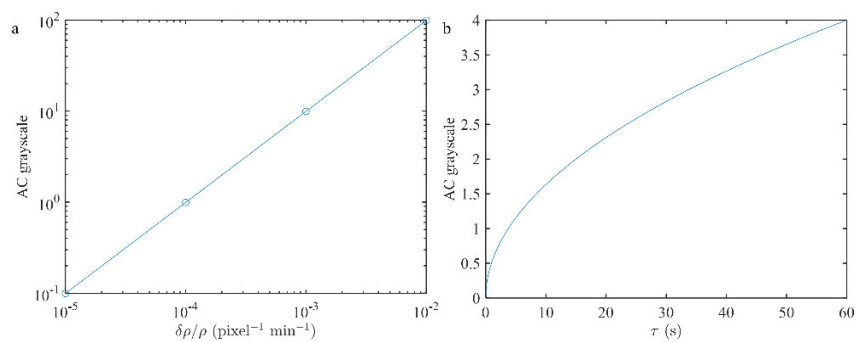

**Supplementary Fig. 1: Frequency domain signal vs. electron density variation and variation duration.** **a.** Frequency-domain signal caused by typical $\delta\rho/\rho$ per pixel during one minute; **b.** Frequency-domain signal from one electron variation in a pixel vs. the duration time of the variation. During the calculation, the refractive index of SF10 substrate is 1.723 and that of solution is 1.335. The electric constant of the gold is taken as -10.562+1.277$i$ and the thickness of the gold film is 48 nm.

## Supplementary References

1     Bürgi, L., Knorr, N., Brune, H., Schneider, M. & Kern, K. Two-dimensional electron gas at noble-metal surfaces. *Applied Physics A* **75**, 141-145 (2002).

2     Snoke, D. W. *Solid state physics: Essential concepts.* (Cambridge University Press, 2020).

3     Bard, A. J. & Faulkner, L. R. *Electrochemical Methods: Fundamentals and Applications, 2nd Edition.* (Wiley, 2000).

4     Azzam, R., Bashara, N. & Ballard, S. S. *Ellipsometry and Polarized Light.* (North Holland, 1978).

5     Jin, X. *et al.* in *4th Optics Young Scientist Summit (OYSS 2020)* (2021).

6     Liu, W., Niu, Y., Viana, A. S., Correia, J. P. & Jin, G. Potential Modulation on Total Internal Reflection Ellipsometry. *Analytical Chemistry* (2016).